\documentclass[prl,twocolumn,a4paper,groupedaddress,showpacs,floatfix]{revtex4-1}
\usepackage{graphicx,amsmath,amssymb,amsfonts,dsfont}
\newcommand{\mf}[1]{\boldsymbol{#1}}
\newcommand{\ket}[1]{\ensuremath{|#1\rangle}}
\newcommand{\mc}[1]{\ensuremath{\mathcal{#1}}}
\newcommand{\bra}[1]{\ensuremath{\langle #1 |}}

\newcommand{\mean}[1]{\ensuremath{ \langle #1  \rangle}}

\usepackage{color}

\begin{document}

\title{Three-body bound states in dipole-dipole interacting Rydberg atoms}

\author{Martin Kiffner${}^{1,2}$}
\author{Wenhui Li${}^{1,3}$}
\author{Dieter Jaksch${}^{2,1}$}

\affiliation{Centre for Quantum Technologies, National University of Singapore, 3 Science Drive 2, Singapore 117543${}^1$}
\affiliation{Clarendon Laboratory, University of Oxford, Parks Road, Oxford OX1 3PU, United Kingdom${}^2$}
\affiliation{Department of Physics, National University of Singapore, 117542, Singapore${}^3$}






\begin{abstract}
We show that the dipole-dipole interaction between 
three identical  Rydberg atoms  can give rise to  bound trimer states. The microscopic origin of these states is 
fundamentally different from Efimov physics. 
Two stable trimer configurations  exist where the atoms form the vertices of an equilateral triangle 
in a plane perpendicular to a static electric field. The triangle  edge length typically exceeds  $R\approx 2\,\mu\text{m}$, 
and each configuration is two-fold degenerate due to Kramers' degeneracy. 
The depth of the potential wells and the triangle edge length can be controlled 
by  external parameters.  
We establish the Borromean nature of the trimer states, analyze the quantum dynamics in the potential wells 
and describe methods for their production and detection. 
\end{abstract}

\maketitle

Rydberg atoms~\cite{gallagher:ryd} are ideal candidates for the investigation of few-body quantum phenomena for several reasons. 
First,  their internal and external degrees of freedom can be accurately controlled and manipulated in state-of-the-art experiments. 
This gives rise to theoretically well-understood and tunable 
dipole-dipole (DD) interactions~\cite{saffman:10,beguin:13} between ultra-cold Rydberg atoms. Second, the range of these DD interactions is 
extremely large - it  typically extends to several microns.  This feature allows the study of few-body quantum systems whose constituents 
can be prepared, manipulated and detected individually. 
Several quantum phenomena arising from strong interactions between two Rydberg atoms  were investigated recently.
Examples are given by the Rydberg blockade effect~\cite{jaksch:00,lukin:01,urban:09,gaetan:09} 
and the realization of quantum gates and entanglement~\cite{wilk:10,isenhower:10}. 
Rydberg atoms can form giant diatomic molecules via different 
binding mechanisms between their constituents~\cite{greene:00,bendkowsky:09,boisseau:02,samboy:11,kiffner:12,overstreet:09}. 
Artificial gauge fields induced by the DD interaction~\cite{yao:12,yao:13} 
and acting on the relative motion of two Rydberg atoms were predicted in~\cite{zygelman:12,kiffner:13,kiffner:13b}. 
In few body-physics, systems with three particles often show  qualitatively different features as compared to two 
particles~\cite{pohl:09,wuster:11,yarkony:96,domcke:ci,efimov:70,kraemer:06,pollack:09,roy:13,braaten:06,wang:11,efremov:13}. 
For example, it has been shown~\cite{pohl:09} that  the dipole blockade can be 
broken by adding a third Rydberg atom.
Furthermore, it has been predicted~\cite{wuster:11} that systems of more than two DD 
interacting atoms exhibit conical intersections~\cite{yarkony:96,domcke:ci}, 
which are relevant for photo-chemical processes. 
A paradigm of few-body quantum physics is  the Efimov effect~\cite{efimov:70,kraemer:06,pollack:09,roy:13}.  
Here a short-range resonant two-body interaction between identical bosons 
gives rise to a universal set of bound trimer states~\cite{braaten:06}. 
Recently, it was shown that the Efimov effect persists~\cite{wang:11} even for a resonant 
 long-range DD interaction. 
%

%
\begin{figure}[b!]
\begin{center}
\includegraphics[width=8cm]{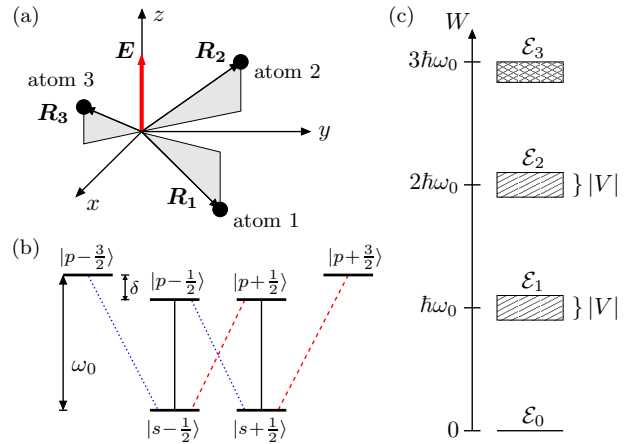}
\end{center}
\caption{\label{fig1}
(Color online) 
(a) System configuration of three DD interacting 
Rydberg atoms. $\mf{R}_{\alpha}$ is the  position  of atom $\alpha$. 
An external electric field $\mf{E}$ is applied in the  $z$ direction. 
(b) Internal level structure of each Rydberg atom.   The Stark shifts 
$\hbar\delta\equiv W_{p \pm1/2}-W_{p \pm3/2}$ of the $\ket{p \pm1/2}$ states  
are negative and $\omega_0$ is the transition frequency  on the $\ket{p\pm3/2}\leftrightarrow\ket{s\pm1/2}$ 
transitions~\cite{kiffner:12}. The dipole transitions  indicated by solid, blue dotted 
and red dashed lines couple to $\pi$, $\sigma^-$ and $\sigma^+$ polarized fields, respectively. 
(c) Level structure of the three-atom state space. 
$\mc{E}_{i}$ contains all states where $i$ atoms are in an $np$ state 
and all other in an $ns$ state. States within  $\mc{E}_1$ and $\mc{E}_2$ 
are coupled by the DD interaction $V$.
}
\end{figure}
%

\begin{figure}[t!]
\begin{center}
\includegraphics[width=8cm]{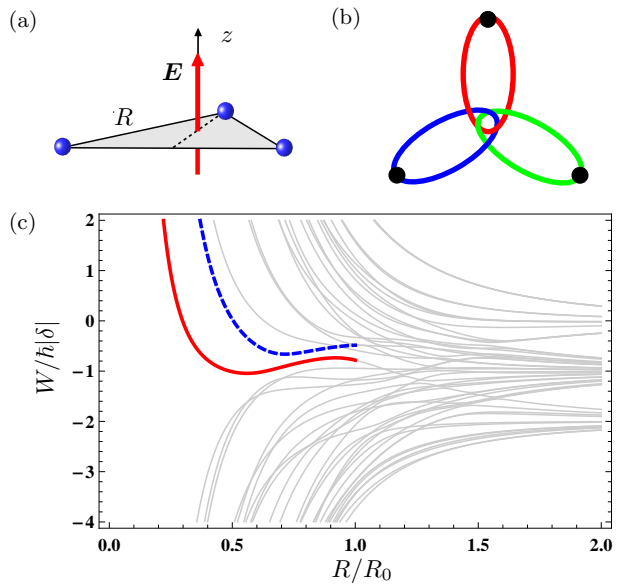}
\end{center}
\caption{\label{fig2}
(Color online) 
(a) Geometry where  the atoms form the vertices of 
an equilateral triangle with  edge length $R$ in the $x - y$ plane. 
The $z$ direction is distinguished by an external electric field $\mf{E}$. 
(b) illustrates the Borromean nature of the trimer configurations. 
(c) All potential curves within the manifold $\mc{E}_2$. The solid red curve 
shows $\epsilon_{\text{t}}$, and the blue dashed curve labels $\epsilon_{\text{s}}$. 
$W+ 2\hbar\omega_0$ represents the absolute energies of the potential curves. 
}
\end{figure}

%
Here we show that  the  DD interaction between three distant Rydberg atoms with non-overlapping electron clouds 
can induce  bound trimer states. 
These states arise from the rich internal level structure of the Rydberg atoms. A crucial 
point is the presence of several dipole transitions in each atom and the interplay between distance-dependent 
DD interactions and Stark-shifted Zeeman states. In stark contrast to the systems described 
in~\cite{efimov:70,kraemer:06,pollack:09,wang:11},  the two-body interaction in our setup cannot 
be described by a single scattering length exceeding all physically relevant length scales. 
The microscopic origin of our states is thus fundamentally different from universal Efimov states. 
The  considered setup is shown in Fig.~\ref{fig1}, and the  level scheme of each atom 
was introduced in the context of Rydberg macrodimers~\cite{kiffner:12}. 
We find  two stable trimer configurations where the atoms form the vertices of an equilateral triangle 
in a plane perpendicular to a static electric field [see Fig.~\ref{fig2}(a)]. 
The triangle  edge length typically exceeds  $R\approx 2\,\mu\text{m}$ and thus the bound states  
occur at extremely long range which is experimentally resolvable. 
We find that our Hamiltonian gives rise to Kramers' degeneracy, and thus the two trimer potential curves are two-fold degenerate. 
The depth of the potential well and the triangle edge length can be controlled 
by the external electric field and the principal quantum number of the Rydberg level. 
We discuss the width and the depth of the potential wells  and describe methods for their production and detection. 
The stable trimer configurations arise from a genuine three-particle effect since two-atom systems are unbound for the 
considered atomic setup. The nature of the trimer bond can thus be illustrated by the Borromean rings shown in Fig.~\ref{fig2}(b):  
If any of the three rings is removed, the remaining two are unbound.

The geometry of the three-atom system  is shown in Fig.~\ref{fig1}(a). 
We denote the position of atom $\alpha$ by $\mf{R}_{\alpha}$, and a 
DC electric field $\mf{E}$ in the $z$ direction  defines the quantization axis. 
The Born-Oppenheimer potential surfaces of this system are determined 
by the Hamiltonian 
\begin{align}
H =  \sum\limits_{\alpha=1}^{3} H_{\text{A}}^{(\alpha)} 
+ V_{13} + V_{23} + V_{12},
\label{H}
\end{align}
where $H_{\text{A}}^{(\alpha)}$ describes the internal levels of atom $\alpha$~\cite{kiffner:12}. 
In each Rydberg atom we consider two angular momentum multiplets as shown in Fig.~\ref{fig1}(b). 
The lower $ns_{1/2}$ states have total angular momentum $J=1/2$, 
and the excited multiplet is comprised of  $np_{3/2}$ states with total angular momentum $J=3/2$. 
We specify the individual atomic states  $\ket{\ell m_j}$ by their orbital angular momentum $\ell$ 
and azimuthal total angular momentum $m_j$. 
The energy difference $\hbar\delta\equiv W_{p \pm1/2}-W_{p \pm3/2}$ denotes the 
Stark shift of the $\ket{p\pm 1/2}$ states with respect to the $\ket{p\pm3/2}$ states. 
The symbol $V_{\alpha\beta}$ in Eq.~(\ref{H}) represents 
the DD interaction~\cite{tannoudji:api} between atoms $\alpha$ and~$\beta$, 
\begin{align}
V_{\alpha\beta}= \frac{1}{4\pi\varepsilon_0 R_{\alpha\beta}^3}[\mf{\hat{d}}^{(\alpha)}\cdot\mf{\hat{d}}^{(\beta)}
-3(\mf{\hat{d}}^{(\alpha)}\cdot\vec{\mf{R}}_{\alpha\beta})(\mf{\hat{d}}^{(\beta)}\cdot\vec{\mf{R}}_{\alpha\beta})], 
\label{vdd}
\end{align}
where $\mf{\hat{d}}^{(\alpha)}$ is the electric dipole-moment operator of atom $\alpha$, 
$\mf{R}_{\alpha\beta} = R_{\alpha}-R_{\beta}$ describes the relative position of atom $\alpha$ with 
respect to atom $\beta$ and 
$\vec{\mf{R}}_{\alpha\beta}=\mf{R}_{\alpha\beta}/R_{\alpha\beta}$ 
is the corresponding  unit vector.

The total state space $\mc{E}$ of the three atoms is spanned by 
the $6^3 = 216$  tensor product states of the individual atoms, 
$\ket{\ell m_j}\otimes\ket{\tilde{\ell} \tilde{m}_j}\otimes\ket{\ell' m'_j}$. 
We divide $\mc{E}$ into four  subspaces $\mc{E}_{i}$ ($i\in\{0,1,2,3\}$), where 
$\mc{E}_{i}$ contains all three-atom states with $i$ atoms in an $np$ state 
and $3-i$ atoms  in an $ns$ state. 
All three-atom states belonging to different subspaces are well separated in 
energy. This is shown in Fig.~\ref{fig1}(c), illustrating that 
states in $\mc{E}_{i}$ are clustered around $i\times \hbar \omega_0$. 
Due to the large energy separation of the subspaces $\mc{E}_{i}$ we neglect any DD induced 
coupling between them  and diagonalize $H$ in Eq.~(\ref{H}) in each subspace independently. 
The DD interaction $V$ in Eq.~(\ref{vdd}) vanishes in $\mc{E}_0$ and $\mc{E}_3$.  On the contrary, 
$V$  has non-zero matrix elements between the $N_1=48$ ($N_2=96$) states in $\mc{E}_{1}$ ($\mc{E}_{2}$).  
Since the bare three-atom states in $\mc{E}_{1}$ ($\mc{E}_{2}$) differ at most 
by $\hbar|\delta|$ ($2\hbar|\delta|$) in energy, they are coupled resonantly by 
the DD interaction $V\propto 1/R^3$ for  interatomic distances $R\le R_0$. 
The characteristic length scale $R_0$ where the strength of 
the  DD interaction equals $\hbar|\delta|$ is
\begin{align}
R_0=[|\mathcal{D}|^2/(4\pi \epsilon_0\hbar |\delta|)]^{1/3},  
\label{r0}
\end{align}
and $|\mathcal{D}|$ is the reduced dipole matrix element of the 
$ns\leftrightarrow np$ transition~\cite{kiffner:12}. The value of $R_0$ can be adjusted 
via the principal quantum number $n$ of the Rydberg level and the DC electric field 
and is typically of the order of several microns  (see  below). 
%
%

%
The Hamiltonian in Eq.~(\ref{H})  is time-reversal invariant~\cite{haake:qsc,supp}. The time-reversal operator is given by  $T = \exp(\imath \pi J_y) \mc{K}$, 
where $J_y$ is the $y$ component of the spin angular momentum operator of the three atoms and $\mc{K}$ denotes complex conjugation with respect 
to the basis states $\ket{\ell m_j}\otimes\ket{\tilde{\ell} \tilde{m}_j}\otimes\ket{\ell' m'_j}$. Since $T^2 = -1$, the time-reversal symmetry gives rise 
to Kramers' degeneracy; every eigenvalue of $H$ in Eq.~(\ref{H}) is (at least) two-fold degenerate.
The main result of this letter is that our system exhibits two 
stable trimer configurations  if the atoms form the vertices of an equilateral 
triangle in the $x-y$ plane [see Fig.~\ref{fig2}(a)]. A first indication for  
three-body bound states is shown in Fig.~\ref{fig2}(c), where all potential curves in the $\mc{E}_2$ manifold 
are shown as a function of the triangle edge length $R$. 
We refer to the potential curve represented by the red solid (blue dashed) line 
as $\epsilon_{\text{t}}$ ($\epsilon_{\text{s}}$).  
The latter two potential curves exhibit a clear minimum, and  are well isolated from all other curves. 
Each potential curve is two-fold degenerate due to Kramers' degeneracy, thus giving rise to four trimer states. 
The minima in  $\epsilon_{\text{t}}$ and $\epsilon_{\text{s}}$ can be explained 
with the same mechanism leading to bound dimer states~\cite{kiffner:12}. As the separation between atom pairs is 
reduced, the DD interaction becomes stronger and eventually couples non-degenerate three-atom states. 
These couplings lead to avoided crossings between repulsive and attractive potential curves and to the formation of potential wells. 

In order to establish that the minima shown in Fig.~\ref{fig2}(c) correspond to bound three-particle states, 
we describe the spatial degrees of freedom in terms of the centre-of-mass coordinate and two relative position vectors $R_{\alpha 3}$  ($\alpha\in\{1,2\}$)~\cite{supp},  
\begin{align}
\mf{R}_{\alpha 3} = R_{\alpha 3} (\sin\theta_{\alpha} \cos\phi_{\alpha},\sin\theta_{\alpha} 
\sin\phi_{\alpha}, \cos\theta_{\alpha}). 
\end{align}
Since the DD interaction in Eq.~(\ref{vdd}) depends only on $R_{1 3}$ and $R_{2 3}$, 
the energy surfaces do not depend on the centre-of-mass coordinates.
In addition, the energies  are independent of the sum $\phi_1 + \phi_2$ due to the azimuthal symmetry of 
the system. We set $\phi_1 + \phi_2 = 0$ such that we have $\phi_1 = \phi/2$ and  
$\phi_2=-\phi/2$, where  $\phi=\phi_1-\phi_2$. 
It follows that all eigenvalues of the Hamiltonian in Eq.~(\ref{H}) are uniquely described in terms of 
five independent variables  $\mf{v} = (R_{13},R_{23},\theta_1,\theta_2,\phi)$. 

\begin{figure}[t!]
\includegraphics[width=8.5cm]{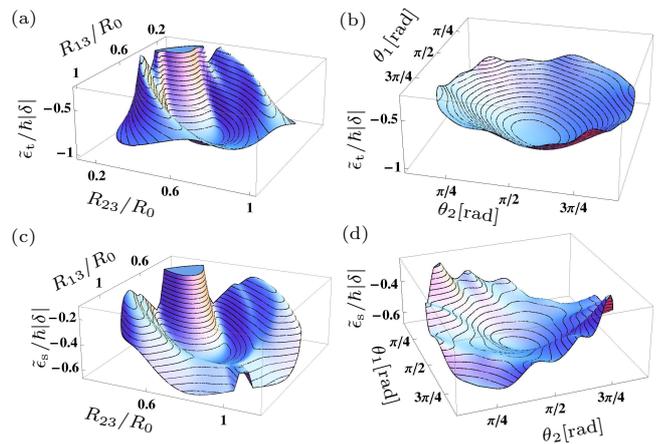}
\caption{\label{fig3}
(Color online) 
(a) Variation of the shifted energy surface $\tilde{\epsilon}_{\text{t}}=\epsilon_{\text{t}}-2\hbar\omega_0$ with 
 $R_{13}$ and $R_{23}$ for $\theta_1=\theta_2=\pi/2$ and $\phi=\pi/3$. 
(b) Dependence of $\tilde{\epsilon}_{\text{t}}$ on $\theta_1$ and $\theta_2$ for $R_{13}=R_{23}=R_{\text{min}}$ and $\phi=\pi/3$. 
(c) and (d) correspond to (a) and (b), respectively,  with $\tilde{\epsilon}_{\text{t}}$ 
replaced by $\tilde{\epsilon}_{\text{s}}=\epsilon_{\text{s}}-2\hbar\omega_0$. 
}
\end{figure}

The potential well $\epsilon_{\text{t}}$ in 
Fig.~\ref{fig2}(c)  has a local minimum 
as a function of the triangle edge length $R$ at $R_{\text{t}} = 0.56 R_0$. 
This minimum corresponds to the parameters 
$\mf{v}_{\text{t}} = (R_{\text{t}},R_{\text{t}},\pi/2,\pi/2,\pi/3)$. 
We evaluate the gradient and the Hessian matrix of $\epsilon_{\text{t}}$ with respect to the independent variables $\mf{v}$ and 
find~\cite{supp}  that $\epsilon_{\text{t}}$ has indeed a local minimum at $\mf{v}_{\text{t}}$. 
For practical purposes it is important to know how deep and how broad the potential minima are around their equilibrium positions. 
The energy surface $\epsilon_{\text{t}}$ is shown in 
Fig.~\ref{fig3}(a) as a function of $R_{13}$ and $R_{23}$ for $\theta_1=\theta_2=\pi/2$ and $\phi=\pi/3$. 
On the other hand, Fig.~\ref{fig3}(b) shows the dependence of $\epsilon_{\text{t}}$ on $\theta_1$ and $\theta_2$ for $R_{13}=R_{23}=R_{\text{t}}$ 
and $\phi=\pi/3$. The width of this potential well about $\theta_1=\theta_2=\pi/2$ is approximately $\pm 45^{\circ}$. 
These results demonstrate that the potential curve $\epsilon_{\text{t}}$ exhibits a relatively broad minimum with respect to 
all variables. 
Similarly, we find~\cite{supp} that  $\epsilon_{\text{s}}$ has a 
local minimum for $\mf{v}_{\text{s}} = (R_{\text{s}},R_{\text{s}},\pi/2,\pi/2,\pi/3)$ with $R_{\text{s}}=0.71\,R_0$. 
The minimum of the potential curve $\epsilon_{\text{s}}$ is more shallow and narrower as compared to 
$\epsilon_{\text{t}}$ [see Figs.~\ref{fig3}(c) and~(d)]. 
The minimal depth of the potential well in  $\epsilon_{\text{t}}$ ($\epsilon_{\text{s}}$) is $0.3\,\hbar|\delta|$ ($0.14\,\hbar|\delta|$). 
%
%

%
We emphasize that 
the bound trimer states arise from a genuine three-particle effect, i.e., they cannot be explained by a pairwise binding of the atoms. 
In order to establish this result, we consider any of the trimer states at the local minimum of the corresponding potential curve 
and calculate  the reduced quantum state $\rho_{\text{dimer}}$ of two atoms.  Since all trimer states are entangled, $\rho_{\text{dimer}}$ 
is a mixed state. We find that $1/3$ of the population of $\rho_{\text{dimer}}$ resides 
in  the $npnp$ subspace which experiences no first-order DD interaction. The remaining $2/3$ of the population 
is found in $nsnp$ states that are DD coupled.  However, all two-atom potential curves are either strongly repulsive or attractive 
for interatomic separations $R_{\text{t}}$ or $R_{\text{s}}$~\cite{kiffner:12,supp},  leading to attraction and subsequent ionization or repulsion on 
timescales much shorter than the Rydberg lifetime~\cite{li:05,amthor:07,viteau:08, park:11b}. It follows that the pairwise interaction cannot 
explain the stability of the trimer configurations which establishes the Borromean nature of the three-particle bound states.  

Next we study the quantum dynamics of the atoms in the trimer configuration by a harmonic approximation of the 
potential well $\epsilon_{\text{t}}$ near its local minimum~\cite{supp}.  Since the 
momenta corresponding to the relative coordinates $\mf{R}_{13}$ and $\mf{R}_{23}$ are coupled in 
the kinetic energy part of the Hamiltonian, the normal modes are not simply given by the 
eigenvalues and eigenvectors of the Hessian matrix  with respect to  the relative coordinates. 
We find that the normal mode with the largest frequency 
$\omega_{\text{s}} = 2.93 \sqrt{\hbar |\delta|/(\mu R_0^2})$ 
is the symmetric stretch mode. There are five different eigenmodes in total, and all frequencies are of the 
same order of magnitude. 
The scissor and asymmetric stretch modes describe oscillations in the $x-y$ plane and 
are degenerate.  The remaining two modes (wagging and twisting) are 
degenerate as well and correspond to oscillations in the $z$ direction. 
The  depth and the spatial extend of the trimer potentials near their local minima can be adjusted by the strength of the DC electric field and 
the principal quantum number $n$ of the Rydberg excitation. The DC electric field is typically of the order of $1\,V/\text{cm}$~\cite{anderson:98} and determines the 
Stark splitting directly, and $\hbar\delta$ and $n$ determine the characteristic length scale $R_0$ in Eq.~(\ref{r0}). 
We note that $\hbar\delta$ and $n$ cannot be chosen independently. In particular, 
the Stark splitting of the $np_{3/2}$ level must be small compared to the $np$ fine structure interval 
$E_{\text{FS}}$. 
In order to give an explicit example, we consider 
Rb or Cs where  $E_{\text{FS}}\sim 0.01n^{-3} E_{\text{h}}$~\cite{gallagher:ryd} 
and $R_0=[E_{\text{h}} n^4 /(3 |\hbar\delta|)]^{1/3} a_0$~\cite{kiffner:12}. 
Here $E_{\text{h}}$ is the Hartree energy and  $a_0$ the Bohr radius. 
A reasonable constraint on $\delta$ is thus given by
$\hbar |\delta|<1.5\times10^{-4}n^{-3} E_{\text{h}}$. Note that 
this choice implies $R_0>(10^{4}n/4.5)^{1/3} n^2 a_0$, ensuring that 
$R_0$  is substantially larger than an individual Rydberg atom. 
For Rb $n=40$ atoms with
$\delta = 2\pi\times10$ MHz we find 
$R_0=4.37\,\mu \text{m}$, $R_{\text{t}}=2.45\,\mu \text{m}$ and  $R_{\text{s}}=3.12\,\mu \text{m}$. 
These parameters result in $\omega_{\text{s}}/(2\pi)\approx 33\,\text{kHz}$ for the largest oscillation 
frequency in the potential well  $\epsilon_{\text{t}}$. 
The lifetime of the trimer states can be calculated from the 300~K radiative lifetimes  of the  $ns$ and 
$np$ Rydberg states~\cite{beterov:09}. 
We find that the lifetime of the trimer state $\ket{\psi_{\text{t}}}$ is  given by $45.5\,\mu\text{s}$ for $n=40$, 
resulting in roughly $1.5$ oscillations in the well before the atoms decay. We emphasize that 
this number can be increased for smaller principal quantum numbers $n$. For example, for Rb $n=30$ atoms with
$\delta = 2\pi\times36$ MHz,  the atoms oscillate 3 times during the lifetime and hence these oscillations 
can in principle be  resolved via spectroscopic techniques. 
The experimental implementation of our scheme can be realized in different ways. Optical potentials allow one 
to prepare three ground state atoms in the triangle configuration with the equilibrium edge length 
$R_{\text{t}}$ or $R_{\text{s}}$. For example, this could be achieved with an optical tweezer setup similar to 
the one described in~\cite{gaetan:09,urban:09}. 
Alternatively,  one could utilize Kagom\'{e} optical lattices~\cite{santos:04} or engineer 
the desired triangular structure via the controlled occupation of specific sites in, e.g., a 
triangular lattice via advanced experimental  techniques~\cite{Weitenberg2011}. 
Subsequently all three atoms are excited in a two-photon process to the $ns$ Rydberg level~\cite{park:11,park:11b,miroshnychenko:10,li:05}. 
These states experience  only a weak van der Waals shift such that the dipole blockade effect is negligible. 
The desired trimer state can then be excited by microwave radiation. Since  states  in the subspace $\mc{E}_2$ 
have no direct dipole matrix element with the $ns$ states in $\mc{E}_0$, the trimer states have to be excited via a 
two-photon process. We find that the trimer states have large dipole matrix elements with states in $\mc{E}_1$, and thus efficient 
excitation processes are possible via tailored microwave fields. 
From recent experiments~\cite{miroshnychenko:10,park:11} we estimate that the two-step excitation can be achieved in less than $1\,\mu \text{s}$. 
Since ultra-cold Rydberg atoms are effectively stationary on this timescale, the atoms remain in their initial configuration during the excitation process. 
A necessary condition for the detection of the trimer states is that their kinetic energy is smaller than the 
depth of the local minima in $\epsilon_{\text{t}}$ and $\epsilon_{\text{s}}$. The minimal  depth for 
the potential curve $\epsilon_{\text{t}}$ ($\epsilon_{\text{s}}$) and for the $n=40$ Rubidium  
parameters above corresponds to a temperature of $144\,\mu\text{K}$ ($67\,\mu\text{K}$). 
This is much higher than typical temperatures of atoms in optical lattices, and 
still higher than the temperature of laser-cooled atoms in optical tweezers~\cite{tuchendler:08}.  
There are several routes towards the detection of atoms in  stable trimer configurations. First, 
they can be detected by spectroscopic methods similar to the suggested procedure for the detection of dimer 
states in~\cite{kiffner:12}. Here the microwave excitation from a trimer state to another state showing no 
DD interaction should exhibit a resonant feature  removed from the atomic transition by the
well depth.
Second, advanced imaging techniques~\cite{schwarzkopf:11,schauss:12} allow one to observe spatially 
resolved patterns of Rydberg excitations.  
While Rydberg atoms in attractive or repulsive three-atom states will quickly leave their initial spatial configuration~\cite{li:05,amthor:07,viteau:08, park:11b}, 
stable trimer 
configurations will stay put. Thus the direct observation of the  triangular configuration of Rydberg atoms after a  time delay following 
their excitation would provide further evidence for the preparation of a trimer state.
In conclusion, we are confident that the trimer states can 
be produced and detected with current experimental techniques. 

In summary, we have shown that three dipole-dipole interacting Rydberg atoms can give rise to bound trimer states. 
Our scheme applies to ultra-cold Rydberg atoms where  experimental techniques offer exquisite control over the external and 
internal atomic degrees of freedom. The trimer states are well separated from other potential curves such that  
the quantum dynamics remains adiabatic.  
Finally, we would like to mention that the opposite regime of non-adiabatic spin dynamics, whose study is beyond the scope of this work, 
might also give rise to novel few body effects. For instance, we find that the spectrum of the system Hamiltonian 
exhibits level repulsion~\cite{haake:qsc}, and hence expect the atomic motion to display chaotic behavior in the semi-classical regime. 
Moreover, in the absence of geometrical symmetries our time-reversal invariant Hamiltonian  belongs to the symplectic symmetry class, 
giving rise to  as yet unobserved quartic level repulsion~\cite{haake:qsc}. 
In conclusion, our system offers fascinating possibilities for future  theoretical and experimental studies with ultra-cold and semi-classical 
Rydberg atoms.

\begin{acknowledgments}
The authors thank  Andreas Buchleitner and Richard Schmidt for helpful discussions and 
acknowledge financial support from the National Research Foundation and the Ministry of Education, Singapore. 
\end{acknowledgments}

\clearpage

\onecolumngrid

\vspace*{0.5cm}
\begin{center}
{\large\bf Supplementary material for:\\[0.3cm]
Three-body bound states in dipole-dipole interacting Rydberg atoms} 
\end{center}

\vspace*{0.5cm}
\twocolumngrid
\section{Time-reversal symmetry}
The action of the time-reversal operator $T = \exp(\imath \pi J_y) \mc{K}$ 
on the three-atom states $\ket{\ell m_j}\otimes\ket{\tilde{\ell} \tilde{m}_j}\otimes\ket{\ell' m'_j}$
changes the sign of the magnetic quantum number in each state, 
\begin{align}
& T\ket{\ell m_j}\otimes\ket{\tilde{\ell} \tilde{m}_j}\otimes\ket{\ell' m'_j} \notag \\
&=\pm \ket{\ell -m_j}\otimes\ket{\tilde{\ell} -\tilde{m}_j}\otimes\ket{\ell'- m'_j}.
\end{align}
Since the atomic level scheme in Fig.~1(a) of the manuscript is symmetric with respect to a  sign change of the magnetic 
quantum number, $T$ commutes with the atomic Hamiltonian,  
\begin{align}
\left[T,\sum\limits_{\alpha=1}^{3} H_{\text{A}}^{(\alpha)} \right] =0 .
\label{THA}
\end{align}
We note that the commutator in Eq.~(\ref{THA}) 
will be different from zero  if the level scheme is made asymmetric~\cite{kiffner:13bb} 
by magnetic or additional AC electric fields. 
It follows that  the time-reversal symmetry of the Hamiltonian $H$ can be broken by external fields. 

In order to evaluate the transformation behavior of the dipole-dipole terms in $H$ in Eq.~(1) of the manuscript, 
we consider the electric dipole-moment operator of one atom and find
\begin{align}
T \mf{\hat{d}}^{(\alpha)}T^{-1} =-\mf{\hat{d}}^{(\alpha)}.
\label{tomin}
\end{align}
Each term in the dipole-dipole Hamiltonian contains a product of two dipole operators, and hence we have 
\begin{align}
T (V_{13} + V_{23} + V_{12})T^{-1} =(V_{13} + V_{23} + V_{12}),
\label{TV}
\end{align}
which implies that $T$ commutes with the dipole-dipole interaction part in $H$. 
Combining Eqs.~(\ref{THA}) and~(\ref{TV}) we obtain 
$[T,H] = 0$, which proves the time-reversal symmetry of the Hamiltonian $H$. 

In the manuscript we describe that we neglect the dipole-dipole coupling between different subspaces 
$\mc{E}_i$ due to the large energy gap between them.  
Here we point out that this approximation leaves the time-reversal symmetry intact, i.e., the approximated Hamiltonian 
is (exactly) time-reversal invariant. In order to show this, 
we write the dipole operator of one atom as 
\begin{align}
\mf{\hat{d}}^{(\alpha)} = \mf{\hat{d}}_+^{(\alpha)} + \mf{\hat{d}}_-^{(\alpha)}, 
\end{align}
where 
\begin{align}
\mf{\hat{d}}_+^{(\alpha)} = \sum\limits_{i=1}^6 \mf{d}_i\, S_{i+}^{(\alpha)}
\end{align}
is the rising part of the dipole operator,  $\mf{d}_i$ and $S_{i+}^{(\alpha)}$ are the dipole moments and atomic rising operators of the 
six dipole transitions indicated in  Fig.~1(b) of the manuscript, respectively, and 
\begin{align}
\mf{\hat{d}}_-^{(\alpha)}=\left[\mf{\hat{d}}_+^{(\alpha)}\right]^{\dagger}. 
\end{align}
The omission of the off-resonant couplings amounts to replacing the dipole-dipole coupling in Eq.~(2) of the manuscript by 
\begin{align}
V_{\alpha\beta}= & \frac{1}{4\pi\varepsilon_0 R_{\alpha\beta}^3}\left[
\mf{\hat{d}}_+^{(\alpha)}\cdot\mf{\hat{d}}_-^{(\beta)}
-3(\mf{\hat{d}}_+^{(\alpha)}\cdot\vec{\mf{R}}_{\alpha\beta})(\mf{\hat{d}}_-^{(\beta)}\cdot\vec{\mf{R}}_{\alpha\beta})
\right. \notag \\
& + \left.\mf{\hat{d}}_-^{(\alpha)}\cdot\mf{\hat{d}}_+^{(\beta)}
-3(\mf{\hat{d}}_-^{(\alpha)}\cdot\vec{\mf{R}}_{\alpha\beta})(\mf{\hat{d}}_+^{(\beta)}\cdot\vec{\mf{R}}_{\alpha\beta})\right]. 
\end{align}
Since the  transformation  relation in Eq.~(\ref{tomin}) holds for the rising and lowering parts of the dipole operator 
individually, 
\begin{align}
T \mf{\hat{d}}_{\pm}^{(\alpha)}T^{-1} =-\mf{\hat{d}}_{\pm}^{(\alpha)},
\end{align}
we find that the approximated Hamiltonian is time-reversal invariant. 
%
%
\section{Local Minima of the Trimer configurations }
Here we show  that $\epsilon_{\text{t}}$ has indeed a local minimum at $ \mf{v}_{\text{t}}$ with 
respect to all independent variables. 
In a first step, we verify that all partial derivatives of $\epsilon_{\text{t}}$ vanish at $\mf{v}_{\text{t}}$. 
To this end we employ the Hellmann-Feynman theorem~\cite{feynman:39} and find 
\begin{align}
\partial_i \epsilon_{\text{t}}(\mf{a}) = \bra{\psi_{\text{t}}} 
\left[\partial_i H\right] \ket{\psi_{\text{t}}} , 
\label{grad}
\end{align}
where we introduced the short-hand notation $\partial_i=\partial/(\partial a_i)$ 
and $\ket{\psi_t}$ is an eigenstate corresponding to $\epsilon_{\text{t}}$. 
A necessary condition for a local minimum of $\epsilon_{\text{t}}$  at $ \mf{v}_{\text{t}}$ is 
\begin{align}
\partial_i \epsilon_{\text{t}}(\mf{v}_{\text{t}}) = 0 , 
\label{zero}
\end{align}
which holds within the available numerical accuracy. 
A sufficient criterion can be derived from the Hessian matrix $M_{\text{H}}$ with components 
$[M_{\text{H}}]_{ij} = \partial_i\partial_j \epsilon_{\text{t}}$. 
The second order derivatives of $\epsilon_{\text{t}}$ can be evaluated according to 
\begin{align}
& \partial_i\partial_j \epsilon_{\text{t}}(\mf{a}) =  \bra{\psi_{\text{t}}} 
\left[\partial_i\partial_j H\right] \ket{\psi_{\text{t}}}  \\
 & + \sum\limits_{\epsilon_{\mu}\not=\epsilon_{\text{t}}} \frac{2}{ \epsilon_{t}-\epsilon_{\mu}} 
\text{Re}\left\{ \bra{\psi_{\text{t}}}\left[\partial_i H\right] \ket{\psi_{\mu}}
\bra{\psi_{\mu}}\left[\partial_j H\right] \ket{\psi_{\text{t}}} \right\}, \notag
\end{align}
where the sum runs over all energies $\epsilon_{\mu}\not=\epsilon_{\text{t}}$ and their corresponding eigenstates $\ket{\psi_{\mu}}$.  
We compute $M_{\text{H}}$ at $\mf{v}_{\text{t}}$ and find that all eigenvalues of $M_{\text{H}}(\mf{v}_{\text{t}})$ 
are strictly larger than zero, and hence 
$M_{\text{H}}(\mf{v}_{\text{t}})$ is positive definite. This establishes that $\epsilon_{\text{t}}$ 
exhibits indeed a local minimum with respect to all independent variables. 
Similarly, we find that  $\epsilon_{\text{s}}$ has a local minimum for $R_{\text{s}}=R_1=R_2=0.71\,R_0$, $\theta_1=\theta_2=\pi/2$ and $\phi=\pi/3$. 
%
%
%
\section{Normal Modes}
\begin{figure}[t!]
\begin{center}
\includegraphics[width=7.5cm]{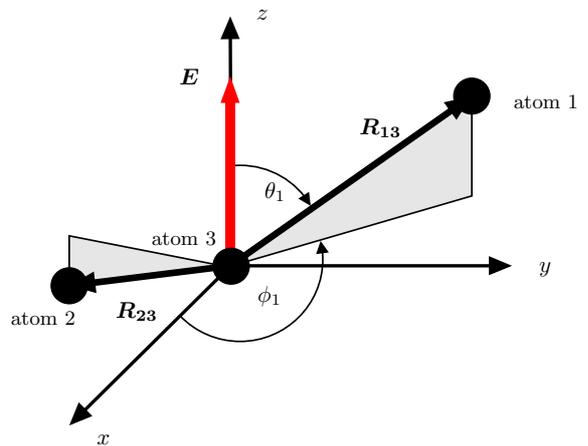}
\end{center}
\caption{\label{fig4}
(Color online) Relative position vectors $\mf{R}_{13}$ and $\mf{R}_{23}$. In Eq.~(4) of the main text, 
$\mf{R}_{\alpha 3}$  ($\alpha\in\{1,2\}$) is expressed in terms of spherical coordinates $R_{\alpha 3}$, $\theta_{\alpha}$ and $\phi_{\alpha}$. 
The angles $\theta_{1}$ and $\phi_{1}$ are indicated in the figure, while $\theta_{2}$ and $\phi_{2}$ were omitted in order to keep the drawing concise. 
}
\end{figure}

Throughout the manuscript we have studied the Born-Oppenheimer surfaces of the Hamiltonian $H$ in Eq.~(1) of the main text 
where all spatial degrees of freedom enter as classical parameters. Here we consider the Hamiltonian of the system 
including the quantized motion of the atoms. For conceptual clarity we label  operator-valued position and momentum variables by a caret. 
The system Hamiltonian $H_{\text{S}}$ of the three Rydberg atoms, including all kinetic energy terms, can be written as 
\begin{align}
H_{\text{S}} = \sum\limits_{\alpha=1}^{3}\frac{\mf{\hat{P}}_{\alpha}^2}{2 m} + H, 
\end{align}
where $\hat{P}_{\alpha}$ is the momentum of atom $\alpha$ with coordinates $\mf{\hat{R}}_{\alpha}$ [see Fig.~1(a) of the manuscript], 
$m$ is the atomic mass and $H$ is the Hamiltonian of the internal degrees of freedom [see Eq.~(1) of the manuscript].
Next we introduce a new set of coordinates and describe the system in terms of the centre-of-mass position and two relative coordinates, 
\begin{align}
& \mf{\hat{R}}_{\text{CM}} = \frac{1}{3}(\mf{\hat{R}}_1 + \mf{\hat{R}}_2 + \mf{\hat{R}}_3), \notag \\
& \mf{\hat{r}}_1 = \mf{\hat{R}}_1-\mf{\hat{R}}_3,\quad \mf{\hat{r}}_2 = \mf{\hat{R}}_2-\mf{\hat{R}}_3.
\label{ctrafo}
\end{align}
Note that the classical counterpart $\mf{r}_1$ ($\mf{r}_2$) of $\mf{\hat{r}}_1$ ($\mf{\hat{r}}_2$) 
equals the relative position vector $\mf{R}_{13}$ ($\mf{R}_{23}$) introduced in the main text and shown in 
Fig.~\ref{fig4}. 
%
The new canonical momenta corresponding to the coordinates in Eq.~(\ref{ctrafo}) are given by 
\begin{align}
& \mf{\hat{P}}_{\text{CM}} = \sum\limits_{\alpha=1}^3\mf{\hat{P}}_{\alpha},\quad 
 \mf{\hat{p}}_1 = \frac{2}{3}(\mf{\hat{P}}_1-\frac{1}{2}\mf{\hat{P}}_2 -\frac{1}{2}\mf{\hat{P}}_3),\\
& \mf{\hat{p}}_2 = \frac{2}{3}(\mf{\hat{P}}_2 -\frac{1}{2}\mf{\hat{P}}_1 -\frac{1}{2}\mf{\hat{P}}_3). 
\end{align}
The new coordinates and momenta obey  canonical commutation relations
\begin{align}
& \left[\hat{R}_{\text{CM}}^{(i)}, \hat{P}_{\text{CM}}^{(j)}\right]=\imath \hbar\delta_{ij}, \\
& \left[\hat{r}_1^{(i)}, \hat{p}_1^{(j)}\right]=\imath \hbar\delta_{ij}, \quad 
\left[\hat{r}_2^{(i)}, \hat{p}_2^{(j)}\right]=\imath \hbar\delta_{ij} ,
\label{commR}
\end{align}
where  superscripts describe Cartesian components of the corresponding vectors and 
$\delta_{ij}$ is the Kronecker delta. All other commutators between coordinate and momentum components vanish. 
In the new coordinates and momenta, the system Hamiltonian can be written as 
\begin{align}
H_{\text{S}} = H_{\text{CM}} + H_{\text{rel}} + H, 
\label{relcoord}
\end{align}
where
\begin{align}
H_{\text{CM}} = \frac{\mf{\hat{P}}_{\text{CM}}^2}{2(3m)}
\end{align}
is the centre-of-mass motion that we omit in the following. 
The kinetic energy of the relative motion is described by 
\begin{align}
H_{\text{rel}} = \frac{1}{2\mu}\mf{\hat{p}}_1^2 + \frac{1}{2\mu}\mf{\hat{p}}_2^2 +\frac{1}{2\mu}\mf{\hat{p}}_1\cdot\mf{\hat{p}}_2 , 
\end{align}
where   $\mu = m/2$ is the  reduced two-body mass. Note that the last term in $H_{\text{rel}}$ mixes the two 
relative momenta. 
We combine all momentum components into one column vector 
\begin{align}
\mf{\hat{p}} =  \left(
\begin{array}{c}
\mf{\hat{p}}_1 \\
\mf{\hat{p}}_2 
\end{array}
\right) 
\end{align}
such that $H_{\text{rel}}$ can be written as 
\begin{align}
H_{\text{rel}} = \frac{1}{2\mu}\mf{\hat{p}}^{\text{t}}  K\mf{\hat{p}}, 
\end{align}
where 
\begin{align}
K = 
\left( 
\begin{array}{cccccc}
1 & 0 & 0 & \frac{1}{2} & 0 & 0 \\
0 & 1 & 0 & 0 & \frac{1}{2}  & 0 \\
0 & 0 & 1 & 0 & 0 & \frac{1}{2}   \\
\frac{1}{2} & 0 & 0 & 1 & 0 & 0 \\
0 & \frac{1}{2} & 0 & 0 & 1 & 0  \\
0 & 0 & \frac{1}{2} & 0 & 0 & 1  
\end{array}
\right) 
\label{Kmat}
\end{align}
and the row vector $\mf{\hat{p}}^{\text{t}}$ is the transpose of $\mf{\hat{p}}$.

Next we determine the vibrational eigenmodes of the system in the trimer state $\ket{\psi_{\text{t}}}$.  
To this end, we combine all coordinates into one column vector 
\begin{align}
\mf{\hat{r}} =  \left(
\begin{array}{c}
\mf{\hat{r}}_1 \\
\mf{\hat{r}}_2 
\end{array}
\right) , 
\end{align}
and its classical counterpart $\mf{r}$ contains the Cartesian coordinates of the relative position vectors $\mf{r}_1$ and $\mf{r}_2$.  
In particular, we denote the Cartesian coordinates of the local minimum of $\mf{\epsilon}_{\text{t}}$ by $\mf{r}_{\text{t}}$. 
We assume adiabatic motion in the internal state $\ket{\psi_{\text{t}}}$ and consider the harmonic 
part of  the potential $\mf{\epsilon}_{\text{t}}$ near its minimum $\mf{r}_{\text{t}}$. 
With these approximations, the potential energy can be written as 
\begin{align}
V_{\text{pot}} = \frac{1}{2}\mf{\hat{x}}^{\text{t}} M \mf{\hat{x}},  
\end{align}
where $\mf{\hat{x}} = \mf{\hat{r}}-\mf{r}_{\text{t}}\mathds{1}$ is the displacement from equilibrium and 
\begin{align}
M_{ij} = \frac{\partial^2}{\partial r^{(i)}\partial r^{(j)}} \left.\epsilon_{\text{t}}\right|_{\mf{r}_{\text{t}}}
\end{align}
are the Cartesian components of the Hessian matrix evaluated at $\mf{r}_{\text{t}}$. 
The eigenmodes  and frequencies are thus determined by the Hamiltonian
\begin{align}
H_{\text{osc}} = \frac{1}{2\mu}\mf{\hat{p}}^{\text{t}}  K \mf{\hat{p}} + \frac{1}{2}\mf{\hat{x}}^{\text{t}} M \mf{\hat{x}} . 
\label{osc}
\end{align}
Here we follow the approach described in~\cite{xiao:09} for the diagonalization of $H_{\text{osc}}$ and note that 
the  matrix $K$ in Eq.~(\ref{Kmat}) is positive definite, and hence a unique Cholesky decomposition of $K$  exists, 
\begin{align}
K = C C^{\text{t}},
\label{kdef}
\end{align}
where $C$ is a real and invertible matrix. 
%
\begin{figure}[t!]
\begin{center}
\includegraphics[width=7.5cm]{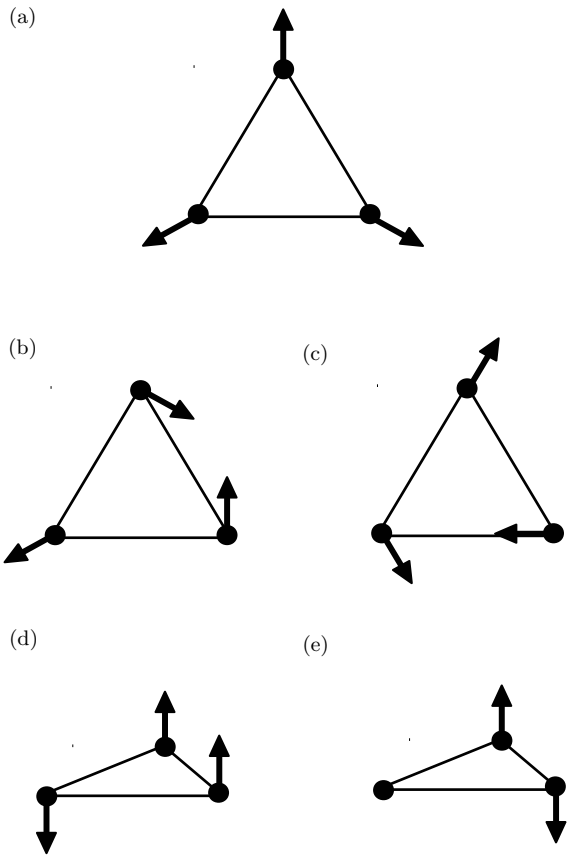}
\end{center}
\caption{\label{fig5}
(Color online) Eigenmodes near the minimum of the potential curve $\epsilon_{\text{t}}$ in the laboratory frame. The arrows 
indicate the displacements of the position vectors $\mf{R}_{\alpha}$.  
The modes can be described as 
(a) symmetric stretch, (b) scissor, (c) asymmetric stretch, (d) wagging and (e) twisting. The modes in (b), (c) and (d), (e) are degenerate.
}
\end{figure}
%
Note that the symmetric $6\times6$ matrix  $M$   
is only positive semi-definite since $\mf{\epsilon}_{\text{t}}$ has not a strict  minimum with respect to all six variables $\mf{r}$. 
In contrast to the five independent variables chosen in the main text, $\mf{r}$ can parametrize atomic configurations that 
leave $\mf{\epsilon}_{\text{t}}$ unchanged due to the azimuthal symmetry of the system. 
Next we introduce the matrix 
\begin{align}
L = C^{\text{t}} M C 
\label{lmat}
\end{align}
 which is real, symmetric and positive semi-definite. It can be diagonalized by an orthogonal 
transformation $S$, 
\begin{align}
S^{\text{t}} L S = \Omega,
\label{odef}
\end{align}
where $\Omega=\mu \,\text{diag}(\omega_1^2,\omega_2^2,\ldots,\omega_6^2)$ is a $6\times6$ diagonal matrix. 
With the definition $A=C S$, Eqs.~(\ref{kdef})-(\ref{odef}) imply 
\begin{align}
A^{-1} K [A^{-1}]^{\text{t}} = \mathds{1}_6,\quad A^{\text{t}} M A =\Omega. 
\end{align}
With the definitions 
\begin{align}
\mf{\hat{q}} = A^{-1} \mf{\hat{x}},\quad \mf{\hat{w}} = A^{\text{t}} \mf{\hat{p}},
\end{align}
we thus find that Eq.~(\ref{osc}) can be written as 
\begin{align}
H_{\text{osc}} = \sum\limits_{i=1}^6\left( \frac{1}{2\mu} \left[\hat{w}^{(i)}\right]^2 + \frac{1}{2}\mu \omega_i^2 \left[\hat{q}^{(i)}\right]^2\right).
\end{align}
The commutation relations in Eq.~(\ref{commR})  and $A A^{-1}= A^{-1} A=\mathds{1}_6$ ensure that 
$\mf{\hat{q}}$ and $\mf{\hat{w}}$ obey canonical commutation relations,
\begin{align}
\left[\hat{q}^{(i)}, \hat{w}^{(j)}\right]=\imath \hbar\delta_{ij}, \quad \left[\hat{q}^{(i)}, \hat{q}^{(j)}\right]=0, \quad 
\left[\hat{w}^{(i)}, \hat{w}^{(j)} \right]= 0 . 
\end{align}
Furthermore,  Eqs.~(\ref{kdef})-(\ref{odef}) allow us to derive the relation
\begin{align}
A^{-1} K M A = \Omega. 
\end{align}
It follows that the normal frequencies are determined by the eigenvalues of the matrix $K M$, and 
the eigenmodes are the corresponding eigenvectors which are the column vectors of $A$.

We find that the normal frequencies of the potential curve $\epsilon_{\text{t}}$ near its local minimum are given by 
\begin{align}
& \omega_1 = 2.93 \times\omega_{\text{vib}},&&  \omega_2 =  \omega_3=1.82\times \omega_{\text{vib}}, \notag \\
& \omega_4 =  \omega_5 = 1.10\times \omega_{\text{vib}},&& \omega_6=0,
\end{align}
where 
\begin{align}
\omega_{\text{vib}} = \sqrt{\frac{\hbar |\delta|}{\mu R_0^2}} . 
\end{align}
In order to visualize the corresponding eigenmodes, we calculate the eigenvectors and 
employ the inverse relations of Eq.~(\ref{ctrafo}) to obtain the displacements of the atoms in the laboratory frame. 
The largest frequency $\omega_1=\omega_{\text{s}}$ belongs to the symmetric stretch mode shown in Fig.~\ref{fig5}(a). The 
doubly degenerate frequency $\omega_2 =  \omega_3$ corresponds to the scissor and asymmetric stretch modes shown in 
Figs.~\ref{fig5}(b) and~(c), respectively. The two degenerate modes  in Figs.~\ref{fig5}(d) and~(e) oscillate with 
frequency $\omega_4 =  \omega_5$ and describe wagging and twisting, respectively. Finally, the last mode with frequency $\omega_6=0$ 
corresponds to a circular motion of the three atoms in the $x-y$ plane and around their centre-of-mass.
A crucial assumption of our approach is that the dynamics in the trimer states remains adiabatic. Physically this is a reasonable 
assumption because the trimer states are well-separated in energy from the remaining states, see Fig.~2(c) of the main text. 
In addition, we have carried out semi-classical calculations in order to verify this assumption explicitly. To this end, we 
start from the full Hamiltonian $H_{\text{S}}$  (without $H_{\text{CM}}$) in Eq.~(\ref{relcoord}) and derive a set of coupled 
equations for the mean values $\mean{\mf{\hat{r}}}$ and $\mean{\mf{\hat{p}}}$~\cite{tannoudji:lc}, 
\begin{align}
\partial_t \mean{\mf{\hat{r}}} & =\frac{1}{\mu} K \mean{\mf{\hat{p}}}, \notag \\
\partial_t \mean{\mf{\hat{p}}} & =- \mean{ \nabla_{\mf{r}}H}, \notag \\
\imath \hbar \partial_t \ket{\psi} & = H(\mean{\mf{\hat{r}}})\ket{\psi},
\label{semicl}
\end{align}
where $\nabla_{\mf{r}} = (\partial/\partial r^{(1)},\ldots,\partial/\partial r^{(6)})^{\text{t}}$ and $\ket{\psi}$ denotes the 
quantum state of the internal three-atom states. At $t=0$ we assume that the system is prepared in 
the trimer state $\ket{\psi_{\text{t}}}$ near the potential minimum and with initial momentum $\mean{\mf{\hat{p}}}_0$. Numerical integration of Eq.~(\ref{semicl}) 
yields $\ket{\psi(t)}$, and the overlap with the trimer state $\ket{\psi_{\text{t}}}$ at position $\mean{\mf{\hat{r}}}(t)$ is a direct measure of 
the adiabaticity of the evolution. We find that the evolution remains adiabatic to a very good approximation for various parameters and kinetic energies 
corresponding to several oscillation quanta $\hbar \omega_{\text{vib}}$.
%
%
\section{Dimer states}
\begin{figure}[t!]
\begin{center}
\includegraphics[width=7.5cm]{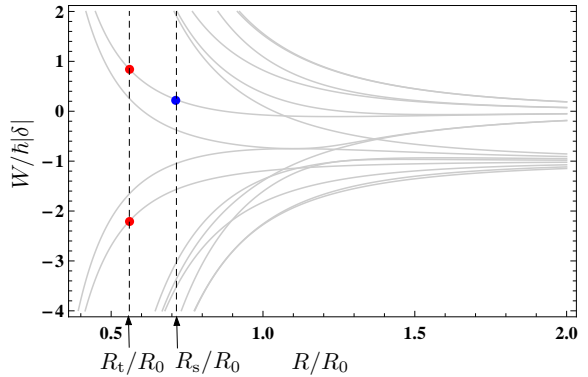}
\end{center}
\caption{\label{fig6}
(Color online) All potential curves in the $nsnp$ subspace of two dipole-dipole coupled Rydberg atoms with the level scheme 
shown in Fig.~1(a) of the manuscript. The atoms are in the $x-y$ plane and $R$ is the atomic separation. 
See text for explanation of the blue and red dots. 
}
\end{figure}

The potential curves of two dipole-dipole interacting Rydberg atoms with the level scheme 
shown in Fig.~1(a) of the manuscript have been investigated in~\cite{kiffner:12b}. There are 16 dipole-dipole coupled  
$nsnp$ states shown in Fig.~\ref{fig6}, where $R$ labels the distance between the atoms and both atoms are located 
in a plane perpendicular to the electric field $\mf{E}$. 
%
%
%
Figure~\ref{fig6} illustrates that all potential curves 
are either strongly repulsive or attractive for $R_{\text{t}}=0.56\, R_0$ and $R_{\text{s}}=0.71\,R_0$, which 
establishes the Borromean nature of the trimer states. In general, the reduced density matrix $\rho_{\text{dimer}}$ of 
two atoms has an overlap with all 16 $nsnp$ states. The states corresponding to the potential curves indicated by the red dots 
contribute the most to $\rho_{\text{dimer}}$ (approximately 16\% each) for the trimer state $\ket{\psi_{\text{t}}}$. 
The state corresponding to the potential curve indicated by the blue dot 
contributes the most to $\rho_{\text{dimer}}$ (approximately 21\% ) for the trimer state $\ket{\psi_{\text{s}}}$.
\end{document}